# Density effect on critical current density and flux pinning properties of polycrystalline SmFeAsO$_{1-x}$F$_x$ superconductor

Y Ding[1,2], Y Sun[1], J C Zhuang[1], L J Cui[1], Z X Shi[1], M D Sumption[2], M Majoros[2], M A Susner[2], C J Kovacs[2], G Z Li[2], E W Collings[2] and Z A Ren[3]
1. Department of Physics, Southeast University, Nanjing 211189, People's Republic of China
2. Center for Superconducting & Magnetic Materials (CSMM), Department of Materials Science & Engineering, The Ohio State University, USA
3. Institute of Physics and Beijing National Laboratory for Condensed Matter Physics, Chinese Academy of Science, Beijing 100190, People's Republic of China

E-mail: zxshi@seu.edu.cn

**Abstract**

A series of polycrystalline SmFeAs$_{1-x}$O$_x$ bulks was prepared to systematically investigate the influence of sample density on flux pinning properties. Different sample densities were achieved by controlling the pelletizing pressure. The superconducting volume fraction, the critical current densities $J_{cm}$ and the flux pinning force densities $F_p$ were estimated from the magnetization measurements. Experimental results manifest that: (1) the superconducting volume fraction increases with the increasing of sample density. (2) The $J_{cm}$ values have the similar trend except for the sample with very high density may due to different connectivity and pinning mechanism. Moreover, The $J_{cm}(B)$ curve develops a peak effect at approximately the same field at which the high-density sample shows a kink. (3) The $F_p(B)$ curve of the high-density sample shows a low-field peak and a high-field peak at several temperatures, which can be explained by improved intergranular current, while only one peak can be observed in $F_p(B)$ of the low-density samples. Based on the scaling behaviour of flux pinning force densities, the main intragranular pinning is normal point pinning.

# 1. Introduction

The discovery of superconductivity at 26 K in the iron oxypnictide LaFeAs(O, F) by Hosono *et al*. [1] soon afterwards lead to the development of iron-based superconductor families with different crystal structures, generally referred to as "1111" for REFeAs(O, F), "122" for AEFe$_2$As$_2$ [2] and AEFe$_2$Se$_2$ [3] , "111" for LiFeAs [4] and "11" for (Fe(Se, Te)) [5]. Here RE denotes rare earth and AE denotes alkali earth. These superconductors attract great interests because of their multiband feature [6, 7], unconventional pairing symmetry [8, 9], and potential applications. The REFeAs(O, F) superconductors, in which $T_c$ is over 50 K when La is replaced by Sm [10], Gd [11] or Tb [12], show very high upper critical fields $H_{c2}$ of about 300 T [13] and high intragranular critical current densities ($J_c$) over $10^6$ A cm$^{-2}$ (5 K, 0 T) [14]. However, the REFeAs(O, F) materials have short coherence length [15], low carrier density [16, 17], significant evidence for granularity and low intergranular $J_c$ [18,19], which are similar to the case of cuprates. Many efforts were devoted to enhance the sample quality [19-21]. The SmFeAsO$_{1-x}$F$_x$ superconducting wires with high transport $J_c$ were successfully prepared by powder-in-tube (PIT) technique of both *in-situ* [22] and *ex-situ* [23] processes. Global supercurrent flow [18], locally well-connected area [24] and intrinsic strong links [22, 23, 25, 26] were confirmed to exist in this system.

However, the transport critical current density $J_{ct}$, which is one of the most important characteristic of superconductors, is still very low. $J_{ct}$ is determined by the flux pinning mechanism activated in the sample and also limited by sample connectivity. The amount of strong links may be influenced by the superconducting volume fraction, sample density and intergranular structure. The cracks, rare earth oxides and Fe-As wetting phase on the grain boundaries of $SmFeAsO_{1-x}F_x$ superconductor were investigated and proved to be detrimental to intergranular current flow [25, 27]. So far there are only a few reports [28-30] concerning the flux pinning mechanism in the REFeAs(O, F) superconductors, and no work related to the influence of sample density. In this paper, we prepared a group of polycrystalline $SmFeAsO_{1-x}F_x$ superconductors with different sample densities and comparatively investigated the density effect on superconductivity in detail.

## 2. Experimental

Several polycrystalline $SmFeAsO_{1-x}F_x$ samples with almost same $T_c$ and different densities were prepared by solid state reaction method through controlling the pelletizing pressure. The precursor SmAs was first synthesized by reacting Sm and As chips in a quartz tube under vacuum (0.001 Pa) at 500℃ for 10 h and then 900℃ for 10 h. The mixture of starting materials SmAs, Fe, $Fe_2O_3$ and $FeF_2$ powders with the nominal stoichiometric ratio of $SmFeAsO_{0.8}F_{0.2}$ were ground thoroughly and pressed into pellets. The pelletizing pressure applied varied from 8 MPa to 20 MPa. The pellets were sintered in the evacuated quartz tubes at 1160℃ for 60 hours. In order to obtain a sample with the density close to the theoretical value, a $SmFeAsO_{0.88}F_{0.12}$ bulk was prepared by high-pressure method [31]. Table I lists the samples being studied (A-F) and their densities.

Powder X-ray diffraction (XRD) was performed on a MAX-RC-type diffractometer with Cu-Kα radiation from $2\theta$=20-80°.The XRD patterns indicated that the main phase was $SmFeAsO_{1-x}F_x$ with trace amounts of SmOF and FeAs impurities. Microstructural and compositional investigations were performed using a Scanning Electron Microscope (SEM, Quanta 200) equipped with an Energy Dispersive X-ray Spectroscope (EDX). Figure 1 (a)-(f) are the typical SEM images of sample A-F, respectively. Sample A shows denser structure and better connectivity than the other samples. For sample B-F, rectangular shaped tabular grains with dimension of about 10 μm can be observed. The cracks between grains are clearly seen and the grain conglomerations are separated by large pores. The SEM images also suggest that the microstructure become more porous with decreasing sample density. The EDX analysis indicated that the composition in the bright regions is homogeneous, and the dark regions have a higher concentration of Fe and As.

DC magnetization measurements were performed using the VSM option of a Quantum Design PPMS with magnetic field applied along the longest dimension of the slabs. Magnetic critical current densities $J_{cm}$ at the same normalized temperature for different samples were estimated using the Bean model from the magnetic hysteresis loops (MHLs).

## 3. Results and discussion

*3.1. Density effect on superconducting volume fraction*

The temperature variation of the DC magnetization was measured to determine the superconducting transition temperature, $T_c$, and to estimate the superconducting volume fraction, $V_{sc}$. Figure 2 shows the zero-field-cool (ZFC) and field-cool (FC) $\chi$-$T$ curves of all samples from 5 K to 120 K under 5 mT. Geometric effect was considered by using an effective demagnetization factor proposed by Brandt [32]. The low values of $\chi$ in the normal state suggest that the influences of magnetic impurities are negligible. $T_c$ was determined by the onset of the diamagnetic signal in ZFC curves. The $T_c$ values of all specimens are also summarized in Table I. As can be seen in figure 2, $V_{sc}$ decreases with decreasing sample density. Sample A with the highest density displays a relatively sharp transition with $V_{sc}$ of about 75% at 5 K. The $V_{sc}$ of sample B-F are less than 30% at 5 K due to their porosity. Sample F with the lowest density also has the lowest $V_{sc}$. It is noted that the ZFC $\chi$-$T$ curve of sample F shows a second transition at about 20 K. This behaviour may be caused by the significant electromagnetic granularity in iron oxypnictide superconductors [18, 19, 33]. The sample density may affect the magnetic critical current density, $J_{cm}$, and will be presented below.

*3.2. Density effect on critical current density*

The MHLs were measured up to 14 T with a ramping rate of 130 Oe/sec. Since $T_c$ of these samples are different, in order to make the comparison more reliable, experimental data were collected at several normalized temperature $t=T_m/T_c$ for each sample. Here $T_m$ is the actual temperature at which the MHLs were measured. Assuming that current flows uniformly over the whole sample, $J_{cm}$ was calculated using the extended Bean model [34] from the MHLs, $J_{cm} = \frac{20\Delta M}{a(1-a/3b)}$. Here $\Delta M$ is the width of the magnetization loop, $a$ and $b$ are the length and width of the slab ($a<b$). As-obtained $J_{cm}$ contains both intragranular and intergranular components due to the granular nature, and it may not be determined accurately since the current loop sizes and their distributions are unknown. Therefore, we mainly focus on comparing the $J_{cm}$ field dependence and the flux pinning behaviour.

Figure 3 (a)-(c) show $J_{cm}$ of sample A-F at $t$=0.1, 0.4 and 0.7, respectively. Three features can be observed:

(1) The $J_{cm}$ values of the sample B-F tend to decrease with the decreasing of density. Since the main structural difference is the sample density, the differences of $J_{cm}$ among sample B-F may mainly caused by the intergranular connectivity. Although the magnetization is dominated by the large contribution of the intragranular currents in the high fields [19, 35], several transport measurements suggest that supercurrent carried by intrinsic strong links persist to high fields and high temperatures [22, 23, 25, 26]. On the other hand, the locally well-connected areas [24] may form "local links" with robust superconductive persisting to several tesla [36]. Higher sample density may increase the amount of local links and strong links, leading to better connectivity thus more intergranular current loops. In figure 3 (b), however, sample D has lower density than sample C but has higher $J_{cm}$ values. This may be due to the different grain boundary structures and properties.

(2) The $J_{cm}(B)$ curves of sample B-F show a peak effect at t=0.4 and 0.7 similar to [35]. The peak effect in $J_{cm}(B)$ was also observed in oxygen deficiency $SmFeAsO_{0.85}$ sample prepared by high-pressure method [18] and $(Ba, K)Fe_2As_2$ [37], but was absent for F-doped samples in [28-30]. Several reports suggest that the peak effect can be explained by the collective pinning and creep model [29, 38, 39]. Another possible reason is that the weak linked areas [40] in sample B-F are suppressed or driven normal with the increment of applied fields, and they may act as pinning centers similar to the role of oxygen deficiency in YBCO [41, 42]. Recently, the peak effect of $J_{cm}(B)$ in oxypnictide polycrystal [36] and single crystal [43] was verified to be caused by a first order phase transition from a ordered "elastically pinned" low-field vortex phase to a high-field disordered phase.

(3) The $J_{cm}$ values of sample A at each temperature are higher than B-F in low fields, and then drop below sample B, C and D with the increasing fields, which is surprising because sample A has the highest density. Sample B was selected as the representative to be further compared with sample A. Figure 4 show $J_{cm}$ of A and B at various $t$. It can be clearly observed that $J_{cm}(B)$ of sample A (dots) develops a kink at $t$=0.2-0.7. The location of the kink is close to the $J_{cm}(B)$ peak of sample B (lines). Both the kink and the $J_{cm}(B)$ peak move to lower fields with increasing temperature. It is very likely that large amount of areas in the samples are weak links or local links due to Fe-As wetting phase on the grain boundary [19, 25, 27], and the number of strong links is very few. The well-connected structure of sample A increases the amount of weak links and local links, leading to the higher $V_{sc}$ and $J_{cm}(B)$ than sample B-F in lower fields. The improved $J_{cm}$ in lower fields also modifies the peak effect into the kink. However, $J_{cm}$ values of sample A in high fields is lower than some samples, which is not likely caused by the difference of the upper critical field $H_{c2}$ because $H_{c2}(T)$ are similar between samples prepared by high-pressure [44] and solid state reaction method [45]. The possible reason is that with the increment of fields, $J_{cm}$ of sample A decreases dramatically due to the weak links and local links being progressively switched off. In high fields, the contribution of intragranular current becomes dominant. Thus, the fact that sample A has finer grains comparing with the long time sintered (60 h) sample B-D results in lower intragranular magnetization thus lower overall $J_{cm}$.

*3.3. Density effect on pinning force density*

The pinning force density $F_p=J_cB$ were calculated to investigate the pinning properties. Figure 5 (a)-(c) display $F_p$ versus $B$ of sample A-F at $t$=0.1, 0.4 and 0.7, respectively. For sample B-F, with increasing sample density, the pinning force density values tend to increase, and the peak in $F_p(B)$ moves to higher fields. The $F_p(B)$ curves of sample B-F show a intermediate field peak, which is commonly caused by the bulk pinning and can be explained by the field dependence of $J_c$ and effective pinning barrier. However, $F_p(B)$ peak of sample A develops in lower fields, and a second high-field $F_p(B)$ peak was observed in figure 5 (b). The double-peak behaviour in $F_p(B)$ has not been reported yet in oxypnictide superconductors.

A comparative study between sample A and B may be helpful to understanding the correlation of the kink and the peak effect in $J_{cm}(B)$. $F_p(B)$ curves of A and B at various $t$ were presented in figure 6 in semilogarithmic scale. It is noticed that a valley appears at the same location of the kink in the $J_{cm}(B)$ curve. The two $F_p(B)$ peaks of sample A and one $F_p(B)$ peak of sample B can be observed at $t$=0.4-0.6. At lower temperature, such as $t$=0.1, the main $F_p(B)$ peak of sample A and sample B are expected to appear at higher fields beyond our measurements. For the SmFeAsO$_{0.8}$F$_{0.2}$ single crystal, the location of the $F_p(B)$ peak is approximately 6 T at 20 K, and 2.5 T at 25 K [30]. In the case of sample A, the location of the high-field $F_p(B)$ peak is about 7 T at $t=0.4$ (21.5 K), and 4 T at $t=0.5$ (26.9 K). Therefore we speculate that the high-field $F_p(B)$ peak of sample A is caused by intragranular current. The low-field $F_p(B)$ peak, which is absent in the single crystal, may originate from intergranular current flowing in local links due to the enhanced sample density. The kink in figure 4 may indicate the applied field at which the intergranular component is suppressed and the intragranular component becomes dominated.

*3.4. Scaling behaviour and pinning mechanism*

To further investigate the pinning properties, $F_p(B)$ was normalized to $f_p=F_p/F_{p\text{-max}}$ as a function of $b=B/B_{\max}$. Here $B_{\max}$ is the field at which $F_p$ reaches its maximum, $F_{p\text{-max}}$. $F_p(B)$ of sample A was normalized using the high-field (intragranular) peak to keep consistent. Figure 7 (a) displays the $f_p(b)$ curves of sample A-F at $t=0.4$. Figure 7 (b) displays the $f_p(b)$ curves of sample B-F at $t=0.7$. Both figure 7 (a) and figure 7 (b) manifest that $f_p(b)$ data of all sample scale well onto one master curve. Figure 7 (c) and figure 7 (d) show the scaling behaviour of $f_p(b)$ of sample A and B at different temperatures. The good scaling behaviour indicates one same pinning mechanism in these samples.

The scaling of $f_p$-$b$ for high-$T_c$ superconductors is often analyzed by using three theoretical models [46]:

$$f(b) = 3b^2(1-\frac{2b}{3}) \quad \text{for } \Delta\kappa \text{ pinning,} \tag{1}$$

$$f(b) = \frac{9}{4}b(1-\frac{b}{3})^2 \text{ for normal point pinning,} \tag{2}$$

$$f(b) = \frac{25}{16}\sqrt{b}(1-\frac{b}{5})^2 \text{ for surface pinning,} \tag{3}$$

The theoretical curves of equations (1)-(3) were presented in figure 7 to determine the pinning mechanism. The overall shape of $f_p(b)$ is more like normal point pinning, although data points of sample B-F scale between normal point pinning and $\Delta\kappa$ pinning as shown in figure 7 (a), (b) and (d). We speculated that the intergranular component modifies the scaling behaviour, moving the $F_p(B)$ peak to higher fields. Results above suggest that the intragranular pinning mechanism activated in sample A-F is normal point pinning.

Figure 8 shows $f_p(b)$ of sample A normalized by the low-field peak. Figure 8 (b) is the magnification of figure 8 (a) with $b$ from 0 to 3. As-obtained $f_p(b)$ curves at various temperatures agree well with the model of surface pinning at low reduced fields. At the fields near $B_{max}$, the plots are scattered and located between the curves of surface pinning and normal point pinning. This result is similar to [28]. It is possible that the grain surfaces of the well-connected areas act as pinning centre for intergranular $J_c$ in low fields. It is worth mentioning that the end point of the fitting indicates the breaking of local links and the domination of intragranular component rather than the approaching of upper critical field $H_{c2}$ or irreversible field $H_{irr}$.

However, $f_p(b)$ behaviour in [30] suggest that $J_c$ in single crystal is determined by surface pinning, which is different from our results in figure 7 (c). This may be influenced by the anisotropy of the crystals because the magnetic fields were applied along the $c$-axis in [30]. More works are needed to further clarify the pinning mechanism in the oxypnictide superconductors.

**Conclusions**

In conclusion, our investigation on a series of polycrystalline SmFeAs$_{1-x}$O$_x$ samples with different densities leads to following conclusions: (1) the superconducting volume fraction, the magnetization critical current $J_{cm}$ and the pinning force density are improved with the increasing sample density due to the improvement of local links. (2) A peak effect in $J_{cm}(B)$ of low-density samples was observed. (3) $F_p(B)$ curves in high-density sample show a double peak behaviour. The low-field peak is caused by improved intergranular current due to enhanced sample density. The high-field peak is related to the dominating intragranular current component in the high fields. (4) The intragranular pinning mechanism activated in this system is normal point pinning.

**Acknowledgments**

We are very grateful to Professors Haihu Wen and Yanwei Ma for their help. This work was supported by the Natural Science Foundation of China, the Ministry of Science and Technology of China (973 project: No. 2011CBA00105), Important Project (Grant No. 2011seuzd09), the US Department of Energy, Division of High Energy Physics (Grant No.DE-FG02-95ER40900) and Scientific Research Foundation of Graduate School (Grant No. YBJJ0933) of Southeast University.

TABLE I. Critical temperatures and densities of the samples.

| Sample | $T_c$ (K) | Density (g/cm$^3$) |
|--------|-----------|---------------------|
| A | 53.4 | 7.25 |
| B | 50.8 | 5.45 |
| C | 49.4 | 4.83 |
| D | 47.8 | 4.67 |
| E | 48.1 | 4.41 |
| F | 45.8 | 4.30 |

**Figure Captions**

Figure 1. (a)-(f) are typical secondary electron microscopy images of sample A-F, respectively.

Figure 2. The ZFC and FC $\chi(T)$ curves of sample A-F under 5 mT from 5 K to 120 K.

Figure 3. The field dependence of critical current density $J_{cm}$ of sample A-F at the normalized temperature $t=T_m/T_c$, (a) $t=0.1$, (b) $t=0.4$, (c) $t=0.7$. Here $T_m$ is the actual temperature at which the measurements were performed.

Figure 4. The field dependence of $J_{cm}$ of sample A (dots) and B (lines) up to 14 T at $t=0.1$-$0.7$.

Figure 5. The field dependence of pinning force density $F_p(B)$ of sample A-F at (a) $t=0.1$, (b) $t=0.4$, (c) $t=0.7$.

Figure 6. The field dependence of pinning force density $F_p(B)$ of sample A (dots) and B (lines) at $t=0.1$-$0.7$ in semilogarithmic scale.

Figure 7. The normalized pinning force $f_p$ of sample A-F as a function of $b=B/B_{max}$ at (a) $t=0.4$, (b) $t=0.7$. (c) $f_p(b)$ of sample A normalized by the high-field peak at $t=0.4$-$0.6$. (d) $f_p(b)$ of sample B at $t=0.2$-$0.7$.

Figure 8. (a) $f_p(b)$ of sample A normalized by the low-field peak at $t=0.1$-$0.7$. (b) is the magnifications with $b$ from 0 to 3 of (a).

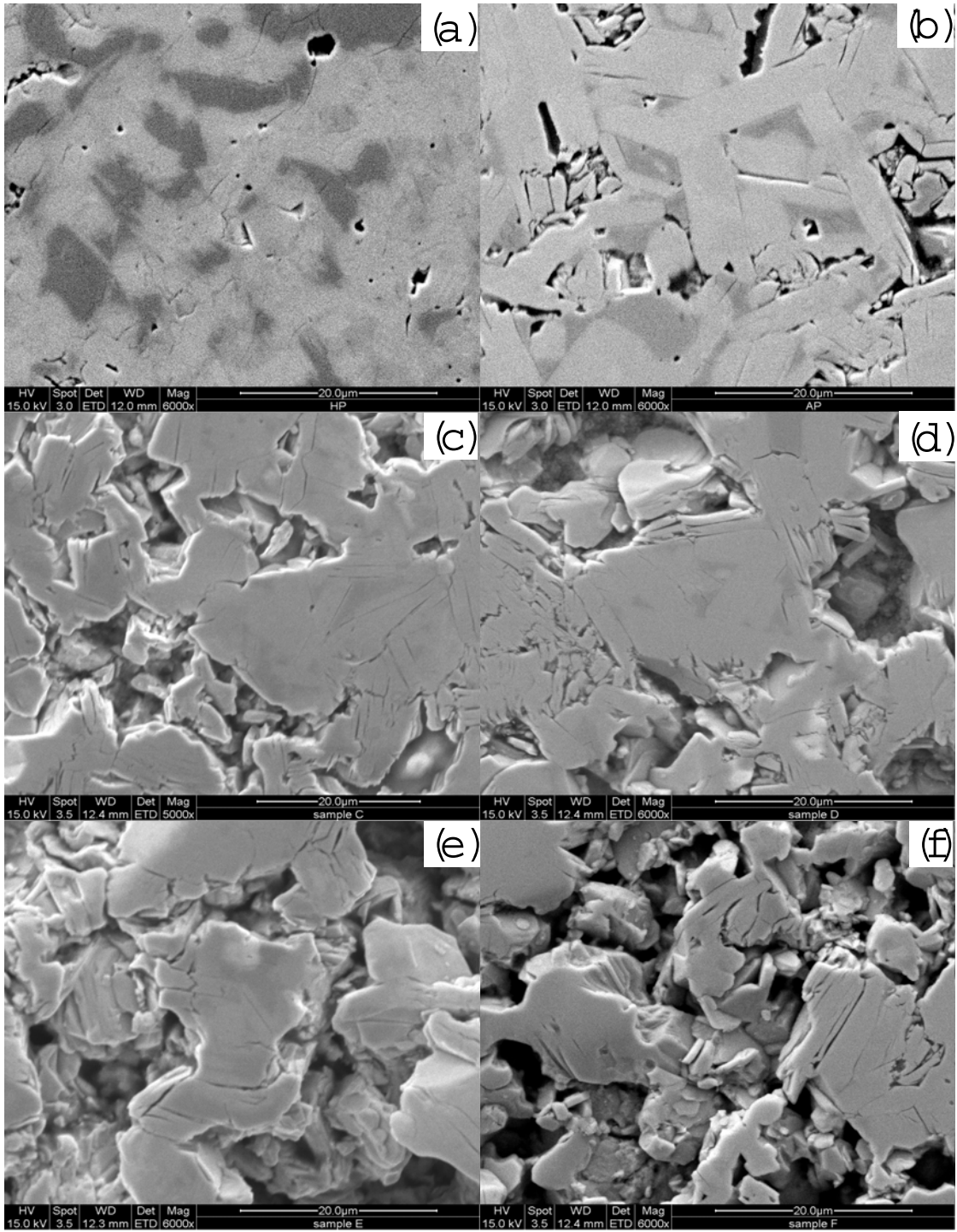

**Figure 1** Y. Ding *et al.*

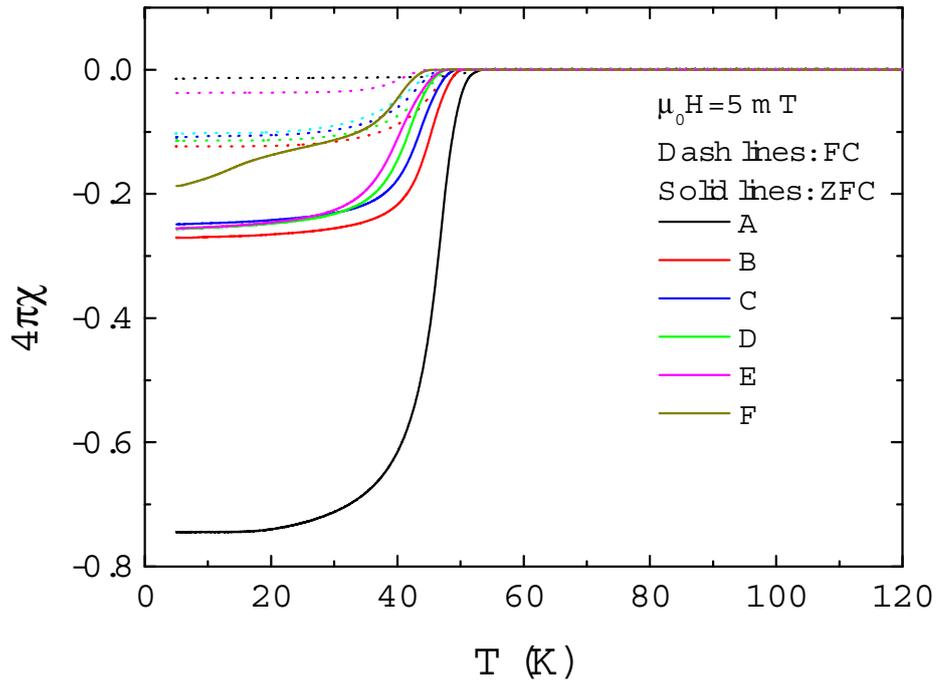

**Figure 2 Y. Ding *et al.***

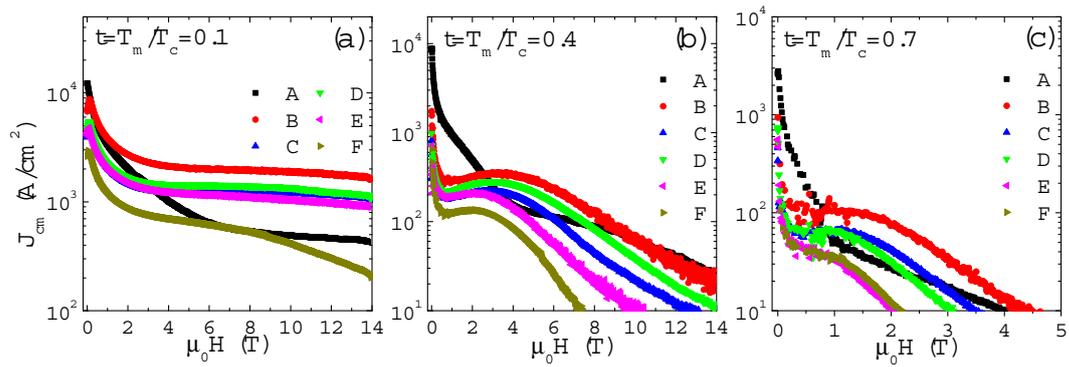

**Figure 3 Y. Ding** *et al.*

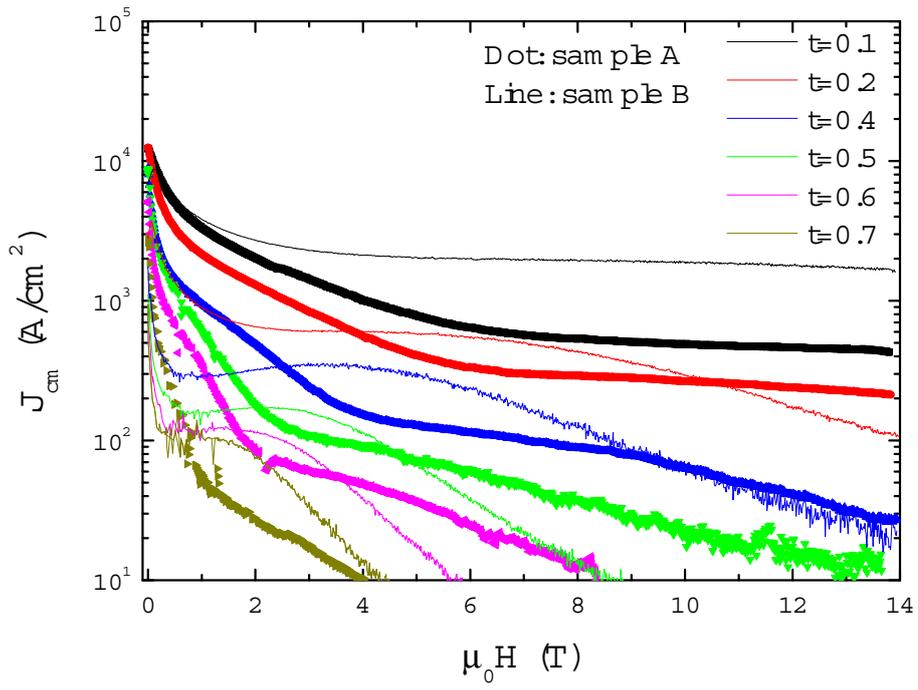

**Figure 4 Y. Ding** *et al.*

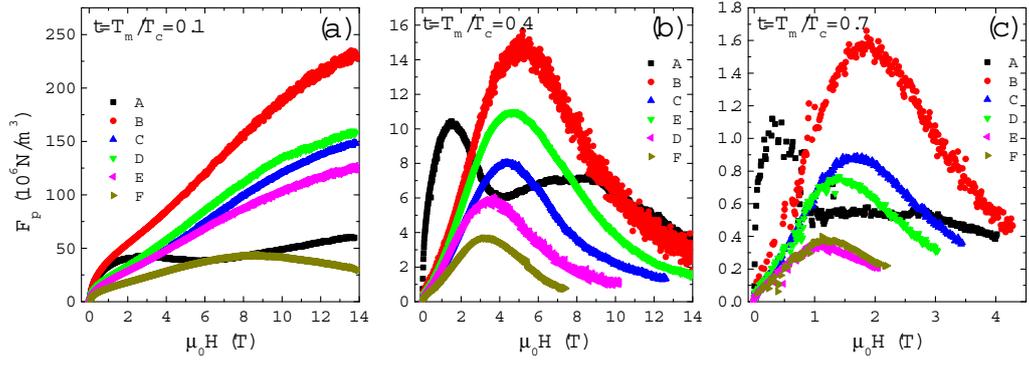

**Figure 5** Y. Ding *et al.*

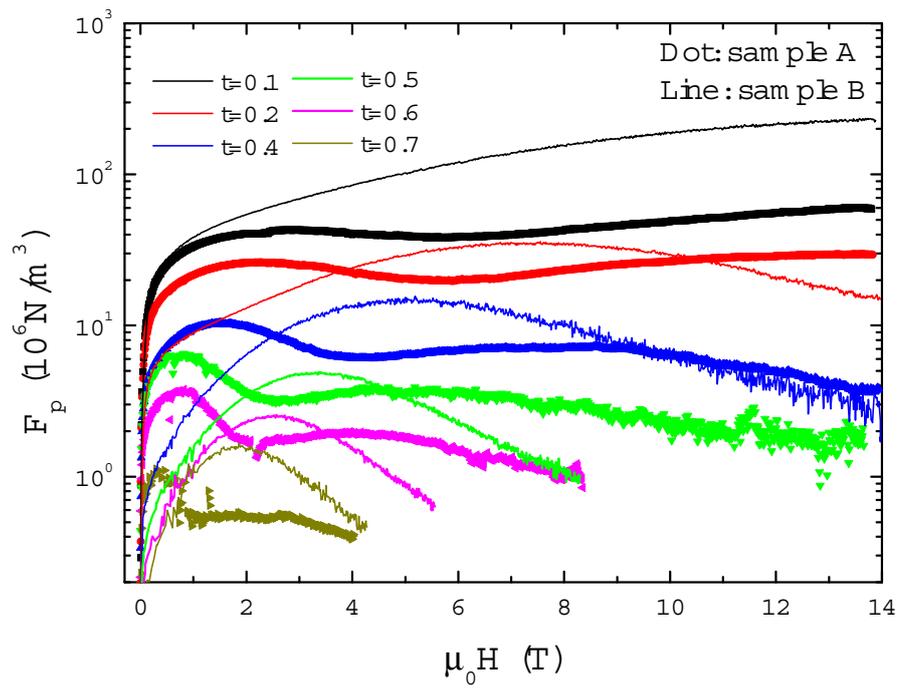

**Figure 6 Y. Ding** *et al.*

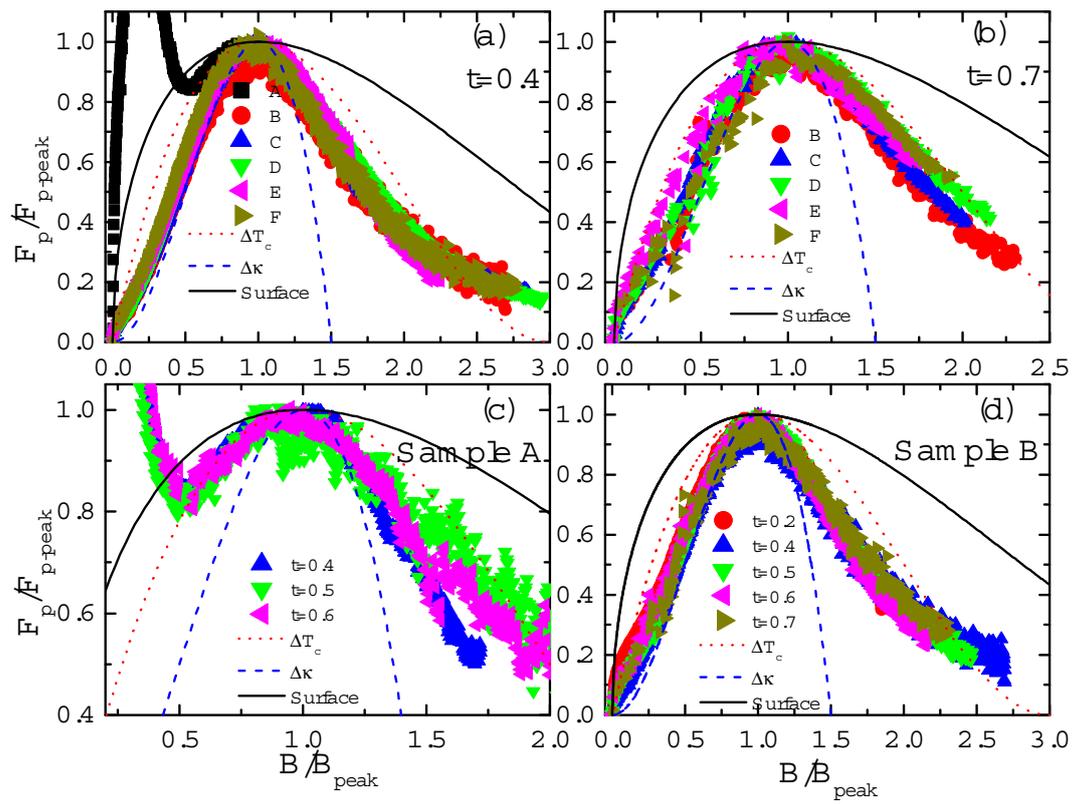

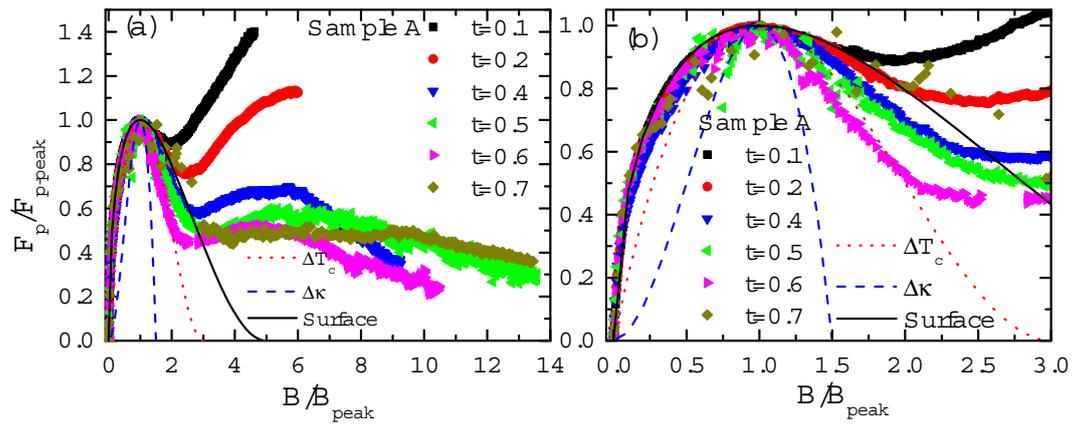

**Figure 8 Y. Ding *et al.***